\documentclass{eas}
\usepackage[pdftex]{graphicx}
\def\be{\begin{eqnarray}}   \def\ee{\end{eqnarray}}
\def\ben{\begin{eqnarray*}} \def\een{\end{eqnarray*}} 

\def\fig#1{Fig.~\ref{f:#1}}

\newcommand{\msun}{\>{\rm M_{\odot}}} 
\TitreGlobal{The Milky Way unravelled by Gaia, Barcelona, Dec 1-5, 2014}
\runningtitle{Chemically tagging the thick disk}

\begin{document}

\title{GALAH survey: chemically tagging the thick disk} 
\author{Joss Bland-Hawthorn}\address{Sydney Institute for Astronomy, University of Sydney, NSW 2006, Australia}
\author{Sanjib Sharma}\address{Sydney Institute for Astronomy, University of Sydney, NSW 2006, Australia}
\author{Ken Freeman}\address{Mount Stromlo Observatory, Australia National University, Woden, ACT 2116, Australia}
\begin{abstract}
The GALAH survey targets one million stars in the southern hemisphere
down to a limiting magnitude of $V=14$
at the Anglo-Australian Telescope. The project aims to measure up to 30 elemental 
abundances and radial velocities ($\approx$ 1 km s$^{-1}$ accuracy) 
for each star at a resolution of $R=28,000$. These elements fall into 8 independent
groups (e.g. $\alpha$, Fe peak, r-process). For
all stars, Gaia will provide distances to 1\% and transverse velocities to 1 km s$^{-1}$ or
better, giving us a 14D set of parameters for each star, i.e. 6D phase space 
and 8D abundance space. There are many scientific applications but here we focus on
the prospect of chemically tagging the thick disk and making a direct measurement
of how stellar migration evolves with cosmic time.
\end{abstract}
\maketitle

% --------------------------------------------------------------------------------
\section{Introduction}
\label{s:intro}
\begin{figure}
  \centering \includegraphics[width=0.85\textwidth]{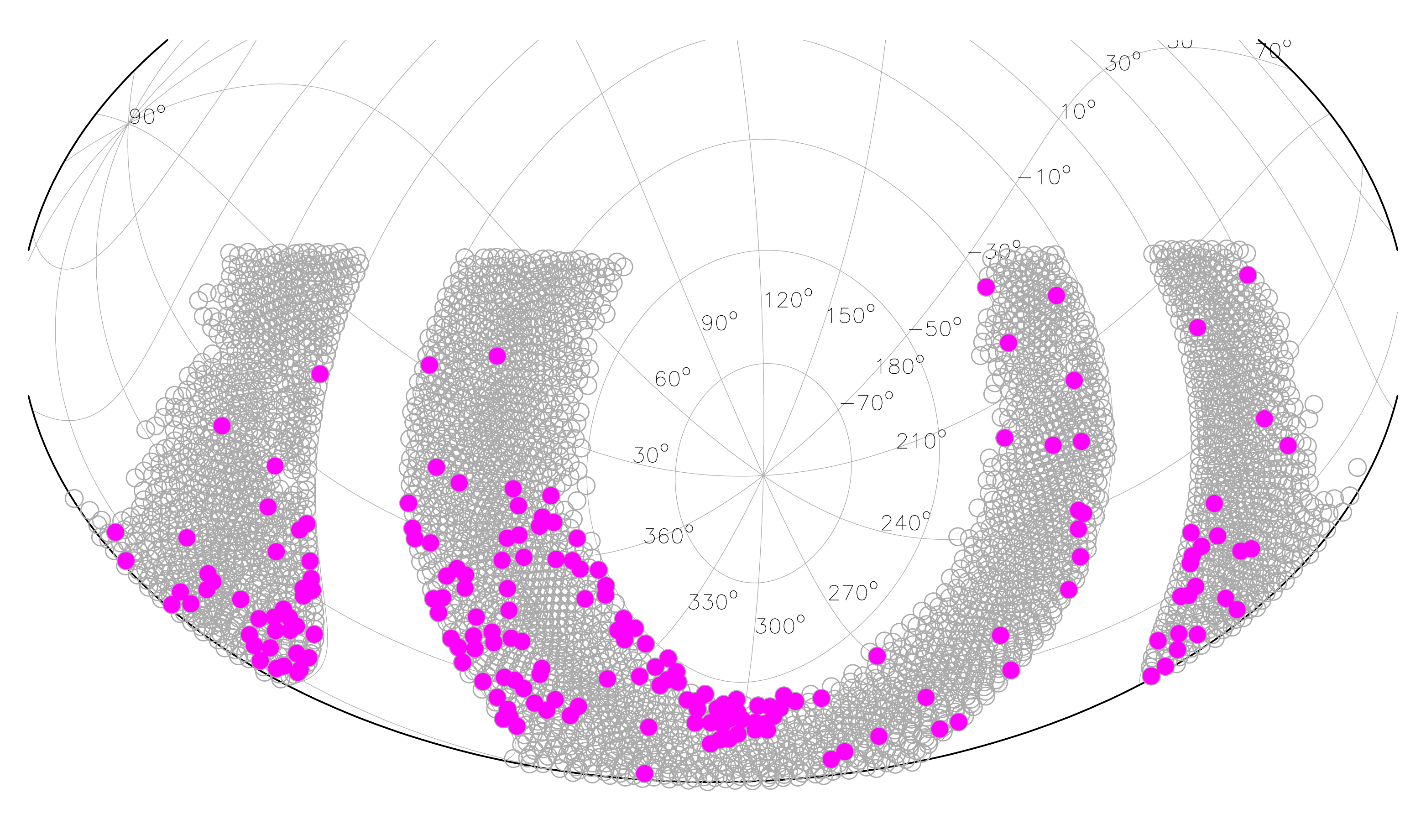}
\caption{Potential GALAH survey fields (4300 in total) across the southern sky 
are shown as open grey circles; the magenta 
circles indicate the fields observed up to November 2014. The full survey
will target about three quarters of the potential fields.
\label{f:f0}}
\end{figure}

The GALAH survey (2014-2019; see Fig.~\ref{f:f0}) 
is the latest major undertaking at the Anglo-Australian Telescope \cite{desilva15}. 
It uses the \$12M HERMES instrument fed by 400 fibres that can be
positioned robotically at the prime focus \cite{sheinis14}. The main
science goal of the project is to chemically `trace' different
Galactic components aided by the stellar kinematics with a particular
emphasis on old stars, i.e. stars that were born (half of the stellar
mass) before $z\sim 1$. In each of these
components, we propose to explore chemical tagging, introduced 
as a tool for reconstructing
information that has been lost over billions years as the Galaxy evolves
\cite{freeman02}. The central
idea is that essentially all stars have been born in homogeneous gas clouds
that have long since dispersed \cite{blandhawthorn10a}. Only a small fraction
remain bound today in the form of star clusters (e.g. open clusters). Thus, in
certain instances, a star's siblings may be identified by a unique chemical
signature in a high dimensional chemical space ($\cal {C}-$space). In principle, we
can hope to reconstruct where some stellar families were born, and how
they have become dispersed throughout the Galaxy.

Arguably, the most interesting target is the thick disk believed to be more 
than 10 Gyr old. Of all components, we predict that the thick disk will
carry the strongest signature of chemical tagging. The thick disk has a very
well defined enhancement in $\alpha$/Fe at all metallicities \cite{bensby14}. 
This is important because it means the stars were
likely born in high pressure, bursty star-forming regions \cite{aalto95}.
Massive star clusters have been observed in high-pressure turbulent disks
at high redshift although their association with the thick disk is uncertain
\cite{lehnert14}.
Direct evidence of enhanced $\alpha$/Fe ratios under starburst conditions comes from
the small fraction of metals detected in x-ray winds from these
environs \cite{martin02}.

This has two consequences for chemical tagging: the initial cluster mass function 
(ICMF) is flatter in starburst environs meaning that there are fewer low
mass clusters, and the extreme clusters tend to be more massive than
found in more quiescent environments. Both of these conspire to make the
tagging signal stronger in the thick disk, at least in our models.
The quiescent outer regions of the thin disk are characterised by 
in situ formation of smaller star clusters \cite{larsen09}.

We now investigate the expected number and size of clusters in the
abundance space ${\cal C}$ for different Galactic components in an 
observational survey like GALAH. Specifically, we investigate as to how  
the ICMF slope $\gamma$ and the maximum cluster mass $m_{\rm max}$  
affects our ability to detect clusters in abundance space.

%%%%%%%%%%%%%%%%%%%%%%%%%%%%%%%%%%%%%
%%%%%%%%%%%%%%%%%%%%%%%%%%%%%%%%%%%%%

\section{Method}
\label{s:method}
The  Galaxy is modelled as consisting of the thin disk\footnote{The low
disk estimate is from the Besan\c{c}on model which we use
for consistency here; our new models presented elsewhere have 
updated most of the Besan\c{c}on parameters.} (2.15$\times
10^{10} \msun$), the thick disk (3.91$\times 10^{9} \msun$), 
the stellar halo (7.6$\times 10^{8} \msun$) and the bulge. 
Depending upon the geometry and the magnitude limits of the survey,
different surveys will sample different fractions of the galaxy.
Let $f_{\rm sample}$ be the fraction of stars of a Galactic population
that is randomly sampled by the survey. Let $M_{\rm pop}$ be the 
actual stellar mass of the population, $m_{*}$ is the mean mass of
stars for a given IMF (here assumed to be 0.42) 
and $f_{\rm mix}$ is the fraction of 
mass of the population that can contribute stars to the survey
volume. Then
\be
f_{\rm sample} & = & \frac{N_{\rm survey}^{\rm pop}}{f_{\rm mix} M_{\rm pop}/m_{*}}
\ee
The factor $f_{\rm mix}$ is the fraction of 
mass of the population that can contribute stars to the survey
volume. If stars are uniformly mixed over the whole galaxy then  
$f_{\rm mix}=1$. More realistically, we assume stars born in 
an annulus of width $\Delta R$ around the Sun such that
\be
f_{\rm mix}=\frac{\int_{R_0-\Delta R/2}^{R_0+\Delta R/2} \Sigma(R)
2\pi R dR}{M_{\rm pop}}.
\ee
In practice, $\Delta R$ is never zero because orbit families develop radial 
excursions during their lifetime. To accommodate radial excursions, we 
assume $f_{\rm mix}$ lies in the range 0.25 to 1.
For GALAH we have $N_{\rm survey}=10^6$ stars. We simulate the selection
function using the Galaxia code \cite{sharma11} and find that 
$24$\% are thick disk stars and $75$\% are from the thin disk; 16\% of stars are within 500 pc,
half are within 1 kpc and 85\% within 2.5 kpc.

\section{The initial cluster mass function}
The stars are assumed to be born in clusters and their 
ICMF is modeled as a power law \cite{elmegreen97}.
The cumulative distribution is given by  
\begin{equation}
\xi(>x|\gamma,x_{\rm min},x_{\rm max})=\left( \frac{x_{\rm max}^{1+\gamma}-x^{1+\gamma}}{x_{\rm
    max}^{1+\gamma}-x_{\rm min}^{1+\gamma}}\right)
\label{equ:clus_mass}
\end{equation}
with $-2.5<\gamma<-1$, and size $x$ is in the range $x_{\rm min}$ 
to $x_{\rm max}$.  
The mean cluster size is then
\begin{eqnarray}
\bar{x} =  \int_{x_{\rm min}}^{x_{\rm max}} x\xi(x)dx 
 =
\left( \frac{1+\gamma}{2+\gamma}\right)\left(\frac{x_{\rm max}^{2+\gamma}-x_{\rm min}^{2+\gamma}}{x_{\rm
    max}^{1+\gamma}-x_{\rm
    min}^{1+\gamma}}\right)
\end{eqnarray}
The full cumulative distribution of the number of clusters above 
a certain size $x$ is given by
\be
N(>x|\gamma,x_{\rm min},x_{\rm max}) & = & \left( \frac{x_{\rm max}^{1+\gamma}-x^{1+\gamma}}{x_{\rm
    max}^{1+\gamma}-x_{\rm min}^{1+\gamma}}\right)\frac{M f_{\rm
  mix}/m_*}{\bar{x}(\gamma,x_{\rm min},m_{\rm max})} \\
& = & \left( \frac{x_{\rm max}^{1+\gamma}-x^{1+\gamma}}{x_{\rm
    max}^{1+\gamma}-x_{\rm min}^{1+\gamma}}\right) \left( \frac{2+\gamma}{1+\gamma}\right)\left(\frac{x_{\rm
    max}^{1+\gamma}-x_{\rm
    min}^{1+\gamma}}{x_{\rm max}^{2+\gamma}-x_{\rm
    min}^{2+\gamma}}\right) M f_{\rm
  mix}/m_* \nonumber
\label{equ:clus_mass}
\ee

\section{Predictions for cluster size distribution }
Given $M_{\rm pop}$, $f_{\rm mix}$, $\gamma$, $x_{\rm min}$, $x_{\rm
  max}$ and $N_{\rm survey}$, the 
number of clusters as function of size $n$ in survey is given by
$N(>n/f_{\rm sample}|\gamma,x_{\rm min},x_{\rm max})$. 
As a concrete example, we predict the number of groups in thick 
disk that can be seen by GALAH. Here $N_{\rm survey}^{\rm pop}=2.3\times 10^5$ stars, 
$M_{\rm pop}=3.9 \times 10^{9}\msun$. 

\begin{figure}
  \centering \includegraphics[width=0.75\textwidth,height=0.65\textwidth,keepaspectratio=false,clip=true,trim=50 130 0 100]{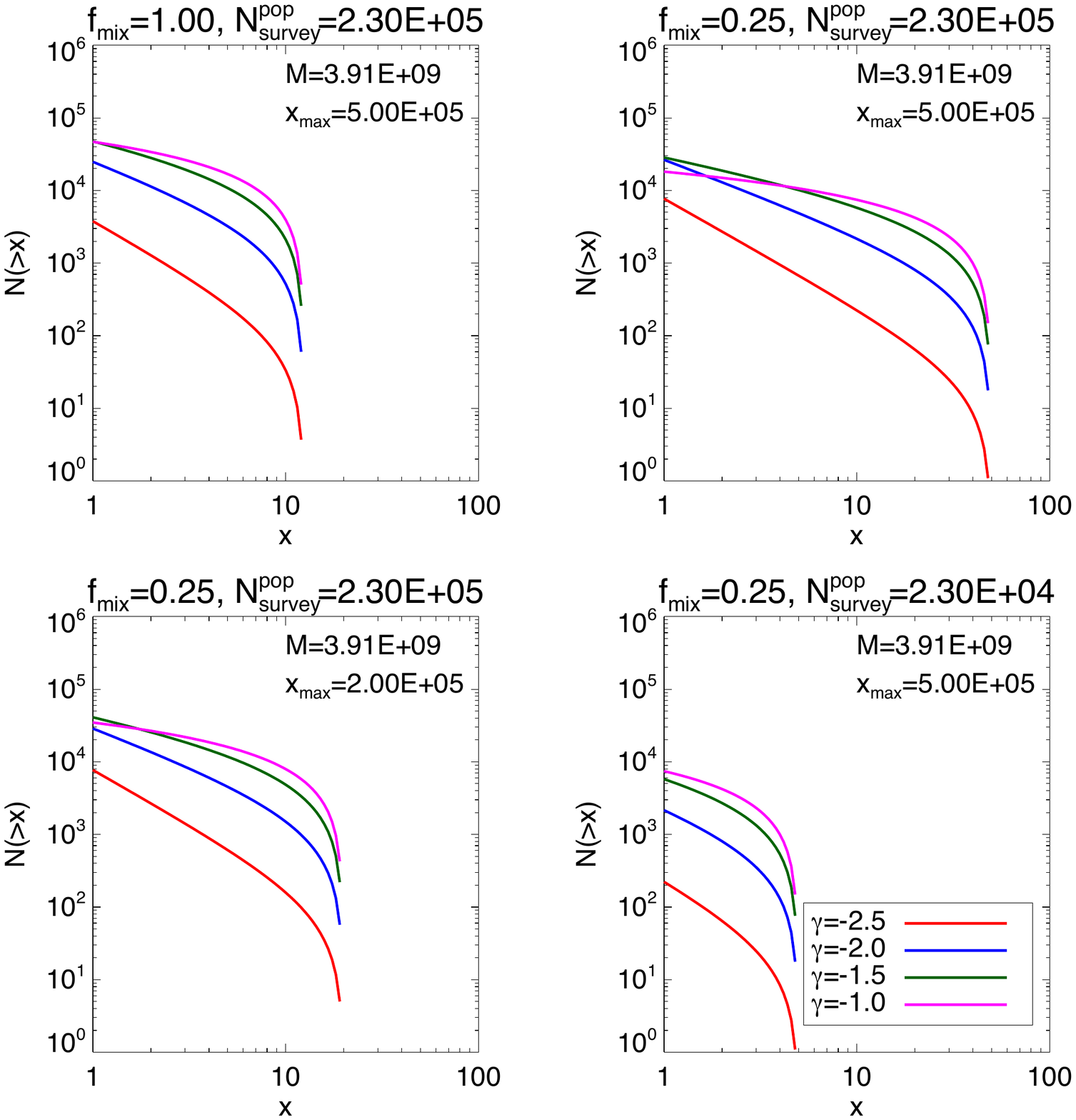}
 % \centering \includegraphics[width=0.65\textwidth,clip=true,trim=50 130 0 100]{figs/thick}
\caption{Cumulative distribution of number of clusters as a function of
cluster size for a simulated thick disk.
\label{f:f1}}
\end{figure}

In \fig{f1}, we show the 
cumulative distribution of clusters as a function of their size 
in the survey. Each panel shows results for four different values of 
$\gamma$. The panels differ in the values of 
$f_{\rm mix}$, $N_{\rm survey}$ and $x_{\rm  max}$.  It can be seen 
that if $f_{\rm mix}$ is large we get fewer big clusters (top two 
panels). If $x_{\rm max}$ is small, we expect fewer big 
clusters. If $\gamma$ is small we expect to see a larger 
number of big clusters. The bottom right panel shows 
that if the number of survey stars is decreased by a factor of 10, 
this dramatically reduces the chances of 
detecting groups with size greater than $\le 10$.

\begin{figure}
  \centering \includegraphics[width=0.75\textwidth,height=0.65\textwidth,keepaspectratio=false,clip=true,trim=50 130 0 100]{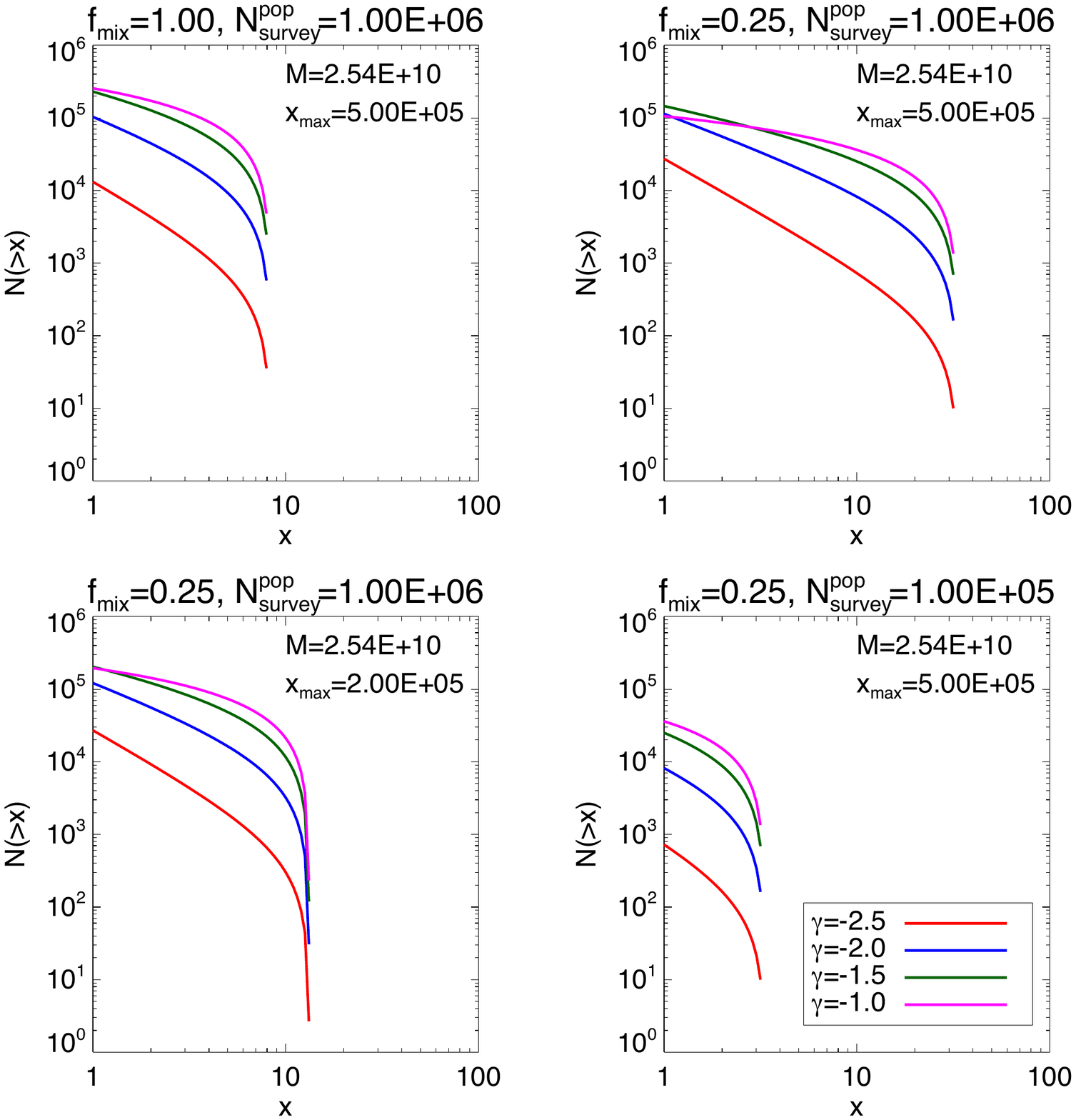}
\caption{Cumulative distribution of number of clusters as a function of
cluster size for all stars, i.e. both thin and thick disk stars.
\label{f:f2}}
\end{figure}

In \fig{f2}, we show the case for all stars. Here $N_{\rm survey}^{\rm pop}=10^6$ stars, 
$M_{\rm pop}=2.55\times 10^{10} M_{\odot}$. The total number of clusters 
increase but the maximum size of clusters is smaller. This is because 
as compared to the thick disk case the ratio 
$N_{\rm survey}^{\rm pop}/M_{\rm pop}$  is smaller.  

%%%%%%%%%%%%%%%%%%%%%%%%%%%%%%%%%%%%%
%%%%%%%%%%%%%%%%%%%%%%%%%%%%%%%%%%%%%

%%%%%%%%%%%%%%%%%%%%%%%%%%%%%%%%%%%%%%
%%%%%%%%%%%%%%%%%%%%%%%%%%%%%%%%%%%%%%

%%%%%%%%%%%%%%%%%%%%%%%%%%%%%%%%%%%%%
%\begin{figure}
%  \centering \includegraphics[width=0.6\textwidth]{thick0}
%\caption{ Cumulative distribution of number of clusters as a function of
%  cluster size. Case shown is for thick disc.
%\label{f:f3}}
%\end{figure}
%%%%%%%%%%%%%%%%%%%%%%%%%%%%%%%%%%%%%

\section{Survey simulations}
\label{s:simulations}
The primary motivation for the GALAH survey is the chemical tagging
experiment described in \S~\ref{s:intro}. Our goal is
to identify debris of disrupted clusters and dwarf galaxies. We assume
here that all of the disrupted objects whose orbits pass through a
$\pm$1 kpc-wide annulus around the Galaxy at the solar circle are
represented within the observable horizon. Simulations
show that a random sample of a million stars with
V $<$ 14 will allow detection of about 20 thick disk dwarfs from each
of about 3000 star formation sites, and about 10 thin disk dwarfs from
each of about 30,000 star formation sites \cite{blandhawthorn04}. Even
a few rare outliers will be enormously valuable for tracking stellar migration
over cosmic time. The specific example of the Solar Family is discussed
elsewhere \cite{blandhawthorn10a} and the first searches have already
started \cite{liu15}.
%Is it possible to detect the debris of about 30,000 different star formation sites, 
%using chemical tagging techniques? Are there enough independent cells in
%${\cal C}$-space to make this possible? The GALAH survey explore
%these issues.

The simulated numbers above depend on the details of the
initial cluster mass function of the disrupted objects
and its mass range. Let us assume the ICMF to be a power law
specified by minimum mass x$_{\rm{\rm min}}$, maximum mass x$_{\rm{\rm max}}$
and power law index $\gamma$. Our expectation is that the thin disc
has a steep ICMF (large $\gamma$) and low x$_{\rm{\rm max}}$ typical of a
quiescent star formation history; the thick disk will have a
shallower ICMF (small $\gamma$) and large x$_{\rm{\rm max}}$
representative of the turbulent high-pressure discs seen at high
redshift. The size of the GALAH survey is selected to probe the slope
$\gamma$ and x$_{\rm{\rm max}}$ of the ICMF.

\begin{figure}
\centering
\includegraphics[width=0.85\textwidth]{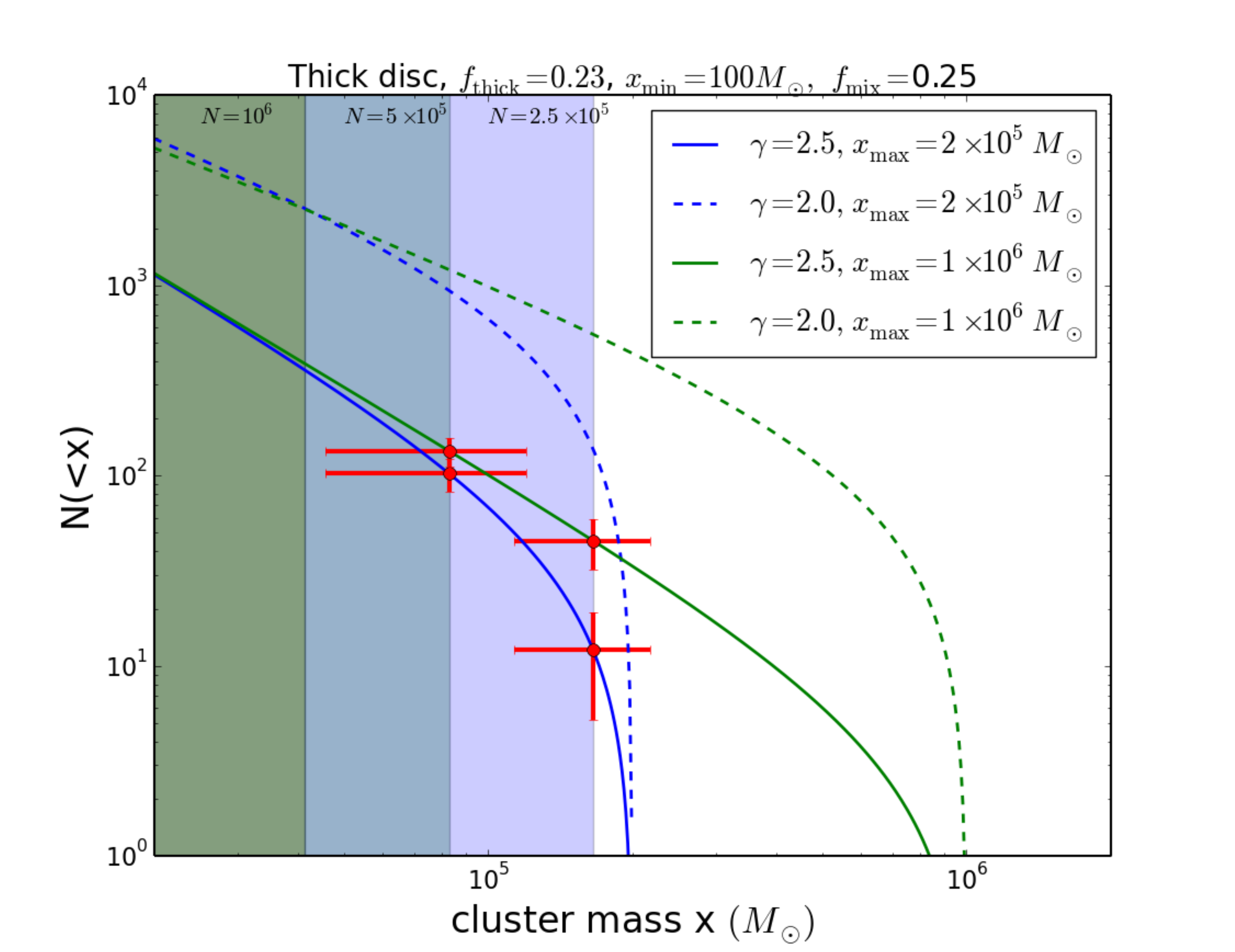}
\caption{The number of clusters recovered from a simulated GALAH
survey as a function of initial cluster mass.  The green, dark blue and light blue shaded regions are the ranges of initial cluster mass accessible to surveys with 10$^6$, $5\times 10^5$ and $2.5\times 10^5$ stars, respectively. The red dots show the cluster masses from which we expect to recover 20 stars (points on left) and 40 stars (points on right) in a million-star survey, and the number of such clusters we would expect to find if 
x$_{\rm{\rm max}}$ is 2 $\times$10$^{5}$ (blue curve) or 1$\times$10$^{6}$ (green curve). The red error bars show the $\pm2\sigma$ uncertainty on the number and size of the recovered groups. The top curves are
for a starburst; the bottom curves are for quiescent star formation.}
\label{f:f3}
\end{figure}

We simulate the GALAH survey based on the Galaxia code \cite{sharma11}
and then applied a simple analytical prescription for generating
stellar clusters. Fig.~\ref{f:f3} shows the number of unique clusters
identified in this simulated survey as a function of initial cluster
mass. Setting the minimum requirement of ten member stars for a
reliable identification of a cluster, we calculate the threshold
cluster mass (i.e. the lowest-mass cluster we expect to recover ten
stars from) as $M_{\rm thresh}$ =10 $\times$ ($M_{\rm pop}$ $\times$
$f_{\rm mix}$)/($f_{\rm pop}$ $\times$ $N$), where $M_{\rm pop}$ is the total mass
of the Galactic population (thin or thick disc),
$N$ is the number of stars in the survey and
$f_{\rm pop}$ is the
fraction of stars in the survey that belong to the population.
The $f_{\rm mix}$ is the fraction of star forming mass of the population that lies
within the survey volume. If there is no mixing $f_{\rm mix}=M'_{\rm
  pop}/M_{\rm pop}$, with $M'_{\rm
  pop}$ being the physical mass of the population enclosed within the
survey volume. If mixing is maximal, i.e., stars born
anywhere within the Galaxy can lie in the survey volume, then 
$f_{\rm mix}=1$.
For the thick disc, $M_{\rm pop}$ =3.9 $\times$
10$^{9} \msun$ and $f_{\rm pop}$=0.236, which means that $M_{\rm thresh}$=4.2
$\times$ 10$^{4} \msun$ for $N$=10$^{6}$ and we assume $f_{\rm mix}$=0.25. Clusters with
initial masses below 4.2 $\times$ 10$^{4}$ are outside the detection
limits of the survey as noted by the green shaded region.  Less
efficient radial mixing brings stars from a smaller number of clusters
into our survey volume, moving all of these thresholds toward lower
mass and making cluster identification easier.

In Fig.~\ref{f:f3}, the red dots show the cluster masses
from which we would expect to recover 20 stars (points on the left)
and 40 stars (points on the right) in a million-star survey, and the
number of such clusters we would expect to find if x$_{\rm{\rm max}}$ is 2
$\times$10$^{5}$ (blue curve) or 1$\times$10$^{6}$
(green curve). The red error bars show 2$\sigma$ Poisson
uncertainty on the number and on the size of the recovered
groups. We see that a smaller survey size would
mean fewer stars per formation site, from a similar number of
formation sites, and severely limit the range of cluster masses over
which we can explore the ICMF.

The above calculations assume that clustering exists in chemical
abundance space and that the clusters are well separated and
observational errors are small enough ($<$0.1 dex) such that they can
be detected by clustering algorithms. In reality, the detectability of
clusters in ${\cal C}$-space depends upon the dimensionality of this space,
the intercluster separation and the observational uncertainties on
abundance measurements. These questions can only be answered with a
large enough data set that has been homogeneously analysed, such as
the GALAH survey data. 

%%%%%%%%%%%%%%%%%%%%%%%%%%%%%%%%%%%%%
%%%%%%%%%%%%%%%%%%%%%%%%%%%%%%%%%%%%%

\section{Discussion}
The argument for a million star survey can be understood most simply
as follows. Let us assume $x_{\rm max}$ be the maximum mass
of a cluster 
of some galactic population of mass $M$. The size of this cluster
in the GALAH survey will then be
\be
n=x_{\rm max} \frac{f_{\rm pop} N_{\rm survey}}{f_{\rm mix}
  M_{\rm pop}}
\ee
For $f_{\rm mix}=0.1$, $f_{\rm pop}=1.0$ and $M_{\rm pop}=2.5 \times 10^{10}
M_{\odot}$, i.e., considering the whole galaxy, we
have
\be
n=4 \frac{x_{\rm max}}{10^4 M_\odot} \frac{N_{\rm survey}}{10^6}.
\ee
So we need $N_{\rm survey}$ to be large so as to observe at least
groups of size 10 or more.

Finally, two factors make it easier to detect groups in thick disc.
First, the $\gamma$ is probably higher for the thick disc (-1.5) compared to the thin
thin disc ($\le -2$). Secondly, the maximum intrinsic size of groups $x_{\rm max}$
is probably also higher for thick disc. Both these effects imply
that we will have more large groups for thick disc. Additionally,
the ratio $f_{\rm pop} N_{\rm survey}/M_{\rm pop}$  is also higher for thick
disc which also helps. This ratio is purely determined by the
selection function and geometry of the survey. 

The question remains as to how well we can effectively isolate thick 
disk stars in the GALAH survey. We will attempt a selection based on
$\alpha/{\rm Fe}$ but also explore kinematic selection, and their
combination. A differential comparison of the ${\cal C}$-spaces for the thick and thin disk
datasets may be the most direct route to confirming that higher levels of 
clumping exist in the thick disk \cite{blandhawthorn10b}. 

Our model is somewhat independent of how the disk formed in that we do not need to specify whether
the star formation was external or internal to the Galaxy.\footnote{For future reference, `exogenous'
and `endogenous' may be useful adjectives for describing a process, e.g. star formation, that is external or internal to a system.} In principle, some part of
the thick disk may have formed through in situ turbulent processes \cite{lehnert14}.
If the thick disk was formed
through accretion, it may be possible to detect a flattened dark-matter component, 
although this would be difficult to separate from the baryonic component and the dark
halo. The process of accretion may indeed have flung open (and globular) clusters into
the halo and the bulge \cite{kruijssen12}, a signature we can look for in the next few years using both
Gaia kinematics and chemical tagging. These will be chemically distinct from star clusters
formed in low-mass dwarf galaxies since most dwarfs have mean metallicities well below [Fe/H]=-1.

% Endogenous vs. exogenous star formation processes.

%%-----------------------------
%%      your bibliography
%%-----------------------------

% ApJ format
%\bibliographystyle{apj}
%\bibliography{refs.bib}

% includes the ENTIRE refs.bib
%\begin{thebibliography}{99}
%\input{refs.bib}
%\end{thebibliography}

\end{document}